\documentstyle[12pt]{article}

\title{{\bf Dirac sea effects in $K^+$ scattering from nuclei}}
\author{ J.C. CAILLON and J. LABARSOUQUE\\
Laboratoire de Physique Th\'eorique$^{\dag}$,
 Universit\'e Bordeaux I\\ rue du
Solarium, 33175 Gradignan Cedex, France}
\date{}
\begin{document}
\begin{titlepage}
\maketitle
\thispagestyle{empty}
\begin{abstract}
 The ratio $R_T$ of $K^+-^{12}C$ to $K^+-d$ cross sections has been calculated
 microscopically using a boson-exchange $KN$ amplitude in which the $\sigma$
and $\omega$ mesons are dressed by the modifications of the Dirac sea in
nuclear matter. In spite of the fact that this dressing leads to a scaling
of the mesons effective mass in nuclear matter, the effect on the $R_T$ ratio
is found to be weak.\\\\
\end{abstract}

LPTB-93-7  \\
E-mail LABARS@FRCPN11 \\
\footnotesize{$\dag$ Unit\'e associ\'ee au CNRS 764}
\end{titlepage}
 \newpage
 These last years, it has been shown that the in-nuclear-medium modifications
 of the $N\overline{N}$ excitations of the Dirac sea might be important
 in the understanding of some phenomena like, for example, the quenching of the
 Coulomb sum rule in electron-nucleus scattering\cite{ks,ji}. \\
 It has also been shown that these modifications lead to an increase of the
 proton radius and to a decrease of the $\omega$ effective mass in nuclear
 matter\cite{ks}, both phenomena often invoked\cite{sk,br,cl} to explain the
 discrepancies between the
 conventional nuclear physics calculations and the experimental data in
 $K^+$-nucleus scattering.\\
\indent The purpose of the present work was to investigate quantitatively in
what extent the $K^+$-nucleus cross sections are sensitive to these
modifications of the Dirac sea.\\
\indent We have analyzed these effects on the ratio $R_T$ of $K^+-^{12}C$ to
$K^+-d$ total cross sections.
\begin{equation}
\label{r}
R_T = \frac{\sigma_{tot}(K^+-^{12}C)}{6 \cdot \sigma_{tot}(K^+-d)}
\end{equation}
As emphasized by many authors, this ratio is less
sensitive to experimental and theoretical uncertainties than, for example,
differential cross sections, and thus more transparent to the underlying
physics. The ratio $R_T$ has been calculated
from $400 MeV/c$ to $900 MeV/c$.\\
\indent In our calculation, we have used the same basic ingredients as in
ref\cite{cl}: the $K^+$-nucleus  optical potential has been built by folding
the density-dependent
$K^+$-nucleon t-matrix by the nuclear densities, these nuclear densities
have been deduced from electron scattering experiments, and the KN amplitude
has been calculated using the Bonn boson exchange model\cite{bo}.
The $K^+$-nucleus total cross-section has been deduced from the forward
$K^+$-nucleus elastic scattering amplitude using the optical theorem.\\
\indent Since in isoscalar nuclear matter the main part of the KN interaction
comes from
$\sigma$ and $\omega$ exchange, we have considered only the dressing of these
two mesons, the remaining part beeing kept as in free space.
Moreover, since the $K^+$-nucleus forward amplitude is dominated
by forward ($q \simeq 0$) KN scattering processes, the modification of the
Dirac sea polarization will be included at the one-loop approximation.\\
\indent In nuclear matter, the $N\overline{N}$ excitations of the Dirac
sea are modified by the surrounding medium.
The main effect of the nuclear medium on the nucleon states
is the modification
of the Dirac spinor which takes the same form as in free space but with an
effective mass:
\begin{equation}
M^*(\rho) = M + \Sigma_S(\rho)
\end{equation}
where $\Sigma_S(\rho)$ is the scalar part of the nucleon self-energy at
density $\rho$. Up to now, this self-energy has not been unambiguously
determined, but it is commonly admitted as the most realistic that, at
saturation, the effective mass would be approximately 15\% smaller than
in free space\cite{ms}. Thus, we have used here a density dependence in this
way:
\begin{equation}
M^*(\rho) = M (1 - 0.15 \frac{\rho}{\rho_0})
\end{equation}
and we have verified that smooth variations from this linear dependence
don't change significantly the results.\\
\indent Since the parameters of the KN interaction have been
determined\cite{bo} in order to reproduce the KN data in free space,  the
polarization of the Dirac sea in free space has already been taken into
account in the KN amplitude, at least in average. Therefore, in nuclear
matter, this free-space polarization of the Dirac sea must be substracted from
the in-medium one. Moreover, in order to eliminate divergent terms, a
renormalization procedure has to be applied. It has been done as
usual\cite{ks,ji,ch} by requiring that, in free space, the three- and
four-$\sigma$ vertices vanish at zero
momenta  and the corrections to the $\sigma$ and $\omega$
masses and wave functions are zero at $q^2 = \mu^2$ ($\mu$ is a
renormalization
scale which disappears here when we substract the polarization in free space).
This leads, for the $\sigma$ channel, to:
\begin{eqnarray}
\label{si}
\Pi_\sigma(q) & = & \frac{3 g_{\sigma N}^2}{2 \pi^2} [3(M^{*^2} - M^2) - 4(M^*
- M) M \nonumber \\  & &- (M^{*^2} - M^2) \int_{0}^{1} ln\frac{M^{*^2}
- x(1 - x)q^2}{M^2} dx \nonumber \\
  & &- \int_{0}^{1} (M^2 - x(1 - x)q^2) ln\frac{M^{*^2} - x(1 - x)q^2}{
M^2 - x(1 - x)q^2} dx]
\end{eqnarray}
and for the $\omega$ channel to:
\begin{equation}
\Pi_\omega^{\mu \nu}(q) = (\frac{q^\mu q^\nu}{q^2} - g^{\mu \nu}) \Pi_\omega
(q)
\end{equation}
with
\begin{equation}
\label{om}
\Pi_\omega(q) = \frac{g_{\omega N}^2 q^2}{\pi^2} \int_{0}^{1} x(1 - x) ln\frac
{M^{*^2} - x(1 - x)q^2}{M^2 - x(1 - x)q^2} dx
\end{equation}
\indent This modification of the Dirac sea polarization produces a displacement
of the poles of the mesons propagators which occur now at $q_0$ such that
$q_0^2 = m_i^2 + \Pi_i(q_0)$ where $i$ stands for $\sigma$ or $\omega$.
If we define an effective mass for the mesons,
$m_i^*(\rho)$ such that $m_i^*(\rho)^2 = q_0^2$, we can see on fig.1 that we
obtain approximately the scaling law derived by Brown and Rho\cite{brr} from
general symmetry arguments:
\begin{equation}
\frac{m_\sigma^*(\rho)}{m_\sigma} \simeq \frac{m_\omega^*(\rho)}{m_\omega}
\simeq \frac{M^*(\rho)}{M}
\end{equation}
As it has already been pointed out by several authors\cite{br,cl}, a
replacement of the
$\sigma$ and $\omega$ mass in the calculation of the KN amplitude by these
density dependent effective masses would lead to an improvement of the
agreement with the experimental data in $K^+$-nucleus scattering (fig.2, curve
b).\\
\indent However, this approximation would be rather questionable here since the
dominant
contributions to the total $K^+$-nucleus cross-section come from forward
($q \simeq 0$) and not from $q \simeq q_0$ KN scattering, and, as we can see
on table 1,
$\Pi_i(0)$ is very different from $\Pi_i(q_0)$.
If we use now, in the mesons propagator, the expressions
obtained above for the polarizations (eq. \ref{si} and \ref{om}), we obtain
a ratio $R_T$ (fig.2, curve c) very close to that calculated with the
free-space KN amplitude (fig.2, curve a). Indeed, the modification of
the polarization being much smaller
in the forward direction than in the $q \simeq q_0$ region, this result can
be easily understood.\\
\indent Thus, in spite of the fact that the modification of the polarization
of the Dirac sea leads
to a scaling of the $\sigma$ and $\omega$ effective mass in nuclear matter,
we have found that its influence on the $K^+$-nucleus cross-section is weak
and that it does not improve appreciably the agreement with experiment for
the $R_T$ ratio.
   
\newpage
\vspace*{4cm}
\begin{center}
\begin{tabular}{|c|c|c|}  \hline
 $i$ & $\Pi_i(0) \: (MeV^2)$ &  $\Pi_i(q_0) \: (MeV^2)$ \\ \hline
  $\sigma$ & $2.24~10^4$ &  $ -1.22~10^5$  \\ \hline
  $\omega$ & $0.$      &   $-2.25~10^5$ \\ \hline
\end{tabular}          \\
\end{center}
{\bf Table 1}: Modification of the vacuum polarization in nuclear matter
(eq. \ref{si} and \ref{om}) at $q=0$
 and $q=q_0$ (pole of the propagator) in the $\sigma$ and $\omega$
 channels at saturation density.
\newpage
\begin{center}
{\bf Figure captions}
\end{center}
\bigskip

 fig.1: Relative variation of the effective mass for nucleons, $\sigma$ and
 $\omega$ mesons with nuclear density.\\

 fig.2: Ratio $R_T$ of the $K^+-^{12}C$ and $K^+-d$ total cross sections
as a function of $p_{lab}$ calculated, curve (a): with the free-space $K^+N$
interaction, curve (b): with the effective mass approximation for the
$\sigma$ and $\omega$ mesons,
 curve (c): with the expressions \ref{si} and \ref{om} for the polarization of
t
 the Dirac sea in the $\sigma$ and $\omega$ channels.
The experimental
points are taken from ref.\cite{x} (circles), from ref.\cite{e} (squares) and
from ref.\cite{p} (triangles).
\end{document}